\documentclass[twocolumn,showpacs,preprintnumbers,amsmath,amssymb]{revtex4}

\usepackage{graphicx}

\begin{document}

\preprint{}

\title{Evidence for Luttinger liquid behavior
 in crossed metallic single-wall nanotubes}

\author{B. Gao,$^1$
A. Komnik,$^2$ R. Egger,$^3$ D.C. Glattli,$^{1,4}$ and
A. Bachtold$^{1}$}

\affiliation{
$^1$~Laboratoire de Physique de la Mati\`{e}re Condens\'{e}e de
l'Ecole Normale Sup\'{e}rieure, 75231 Paris 05, France \\
$^2$~Physikalisches Institut, Albert-Ludwigs-Universit{\"a}t, 79104
Freiburg, Germany
\\ $^3$~Institut f{\"u}r Theoretische Physik,
Heinrich-Heine-Universit{\"a}t, 40225 D{\"u}sseldorf, Germany\\
$^4$~SPEC, CEA Saclay, 91191 Gif-sur-Yvette, France
}

\begin{abstract}
Experimental and theoretical
 results for transport through crossed metallic single-wall
nanotubes are presented. We observe a zero-bias anomaly in one
tube which is suppressed by a current flowing through the other
nanotube.  The phenomenon is shown to be consistent with the
picture of strongly correlated electrons within the Luttinger
liquid model. The most relevant coupling between the nanotubes is
the electrostatic interaction generated via crossing-induced
backscattering processes. Explicit solution of a simplified model
is able to describe qualitatively the observed experimental data
with only one adjustable parameter.
\end{abstract}

\vspace{.3cm} 

\date{ \today}
\maketitle

Single-wall carbon nanotubes (SWNTs) continue to receive a lot of
attention in connection to electronic transport in interacting
one-dimensional (1D) quantum wires. Metallic SWNTs represent a
nearly perfect 1D system, with $\mu$m-long mean free paths
\cite{bachtold1,kong,liang} and only two spin-degenerate transport
channels, where it has been theoretically predicted that electrons
form a Luttinger liquid (LL) rather than a conventional Fermi
liquid phase \cite{egger,kane}. Experimental evidence for LL
behavior in an individual SWNT has been reported in tunneling
\cite{bockrath,yao,postma1} and resonant tunneling measurements
\cite{postma2}, revealing a pronounced suppression in the
tunneling density of states [zero-bias anomaly (ZBA)].  Although
the observed power-law ZBA can be consistently explained by the LL
theory, it is difficult to rule out alternative explanations based
on, e.g., environmental dynamical Coulomb blockade. Furthermore, a
very similar ZBA has been experimentally observed in multi-wall
nanotubes \cite{bachtold2,tarkiainen,yi} although such systems are
known to be disordered multi-channel wires
\cite{egger01,mishchenko}. It is therefore of importance to
clearly identify Luttinger liquid signatures beyond the ZBA for
tunneling into an individual SWNT \cite{bockrath,yao,postma1}.
Following the proposal of Refs.~\cite{komnik1,komnik2}, in this
paper we report experimental evidence in support of the LL picture
from electrical transport through two crossed metallic SWNTs.

Albeit crossed nanotubes have been investigated by other groups
before \cite{fuhrer,jkim,janssen}, so far no transport
measurements for crossed metallic SWNTs have been reported below
room temperature.
 In our experiments, the conductance is measured
first in one tube while the second is left floating. The
conductance decreases as the temperature or the bias is reduced,
in a way very similar to that of tunneling experiments in SWNTs.
Interestingly, the ZBA disappears as the current is increased
through the second tube. Below we discuss the relationship between
these results and LL predictions. The electrostatic coupling
between the tubes is expected to pin the sliding low-energy
excitations (plasmons). This mechanism is enhanced by the
backscattering generated by the mechanical deformation of the
tubes at the crossing. When current is imposed to flow in one
tube, the pinning tends to be suppressed, enhancing the plasmon
sliding and therefore the current in the second tube. Explicit
calculations based  on LL theory
are able to reproduce the measurements rather well,
while several alternative explanations are shown to be unlikely below.

\begin{figure}
\includegraphics{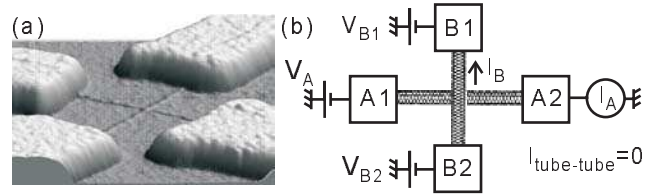}
\caption{\label{fig1} (a) AFM image of a crossed SWNT junction.
The electrode height is $45$~nm. (b) Scheme of the device together
with the measurement setup. The AFM image can not discriminate
which tube lies on top of the other.}
\end{figure}

The SWNT material is synthesized by laser ablation \cite{thess}.
SWNTs are dispersed from a suspension in dichloroethane onto an
oxidized Si wafer. AFM is then used to locate crossed SWNTs with
an apparent height of  $\approx$~1~nm,  presumably corresponding
to individual SWNTs. Next, Cr/Au electrodes are attached using
electron-beam lithography. An example of a device is shown in
Fig.~\ref{fig1}(a).  The separation $L$ between the crossing point
and the electrodes is chosen to be $\simeq$ 300~nm.  For shorter
$L$, undesired finite-size effects may come into play, while for
much longer $L$, the probability is enhanced to find disorder
centers along the SWNTs that complicate the analysis. Devices were
then studied above 20~K, where the thermal length $\hbar
v_{F}/kT$, with Fermi velocity $v_{F}$, remains short compared to
$L$. More than 60 samples have been realized, but we have never
been lucky enough to achieve a device with two crossed metallic
SWNTs and, at the same time, to keep all contact resistances low,
so that Coulomb blockade is negligible. Four times an almost ideal
device has been obtained, where only one of the four contact
resistances was large. Measurements have been carried out on these
four devices, which all gave similar results. A representative set
of measurements on one device is presented next.

In this device, at $T=220$~K,
 the two-point resistances at zero bias of the two SWNTs  [henceforth called
$A$ and $B$] are $R_{A}=$19 k$\Omega$ and $R_{B}=524$ k$\Omega$,
while the four-point resistance of the tube-tube junction is
$R_{X}=277$ k$\Omega$. Other two-point measurements with electrode
$B1$ contacted, see Fig.~\ref{fig1}(b) for the electrode
identification, give also large resistance, suggesting that the
large $R_{B}$ comes from a poor interface between tube $B$ and
electrode $B1$. Note that the two-point measurements are achieved
with the other electrodes left floating. When temperature is
decreased, this large contact resistance induces Coulomb blockade
(CB) oscillations in tube $B$ with zero current for different
regions in the backgate voltage $V_{g}$. In the following, $V_{g}$
is fixed at a broad CB peak.

\begin{figure}
\includegraphics{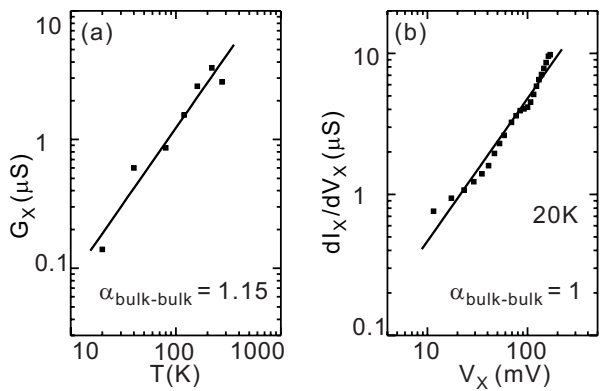}
\caption{\label{fig2} Tunneling measurement on the tube-tube
junction in a four-probe configuration. (a) Linear conductance
$G_{X}(V_{X}=0,T)$. (b) Differential conductance
$dI_{X}/dV_{X}(V_{X},T=20K)$.}
\end{figure}

The device is further characterized by measuring the LL
interaction parameter $g$ \cite{egger,kane} from the tunneling
density of states. Fig.~\ref{fig2} shows the differential
tube-tube tunneling conductance $G_X(V_X, T)=dI_X/dV_X$ measured
in a four-point configuration. Electrons tunnel from the middle of
one SWNT to the middle of the second SWNT (bulk-bulk tunneling).
The double-logarithmic plots of $G_X(V_X,T)$ in Fig.~\ref{fig2}
are in the studied ranges described by a power-law scaling with
slope $\alpha_{\rm bulk-bulk} \simeq 1.1$. Using $\alpha_{\rm
bulk-bulk} =(g^{-1}+g-2)/4$ \cite{egger,kane}, this gives $g
\simeq 0.16$. This value is slightly lower than the generally
reported values $g \simeq 0.2$ for tunneling into a SWNT from a
metal electrode \cite{bockrath,yao,postma1}, reflecting slightly
stronger Coulomb
 interactions among the electrons.  This is presumably due to
different geometries in Refs.~\cite{bockrath,yao,postma1} and in
our device, in particular concerning the size and location of the
connecting electrodes screening part of the interaction.

\begin{figure}
\includegraphics{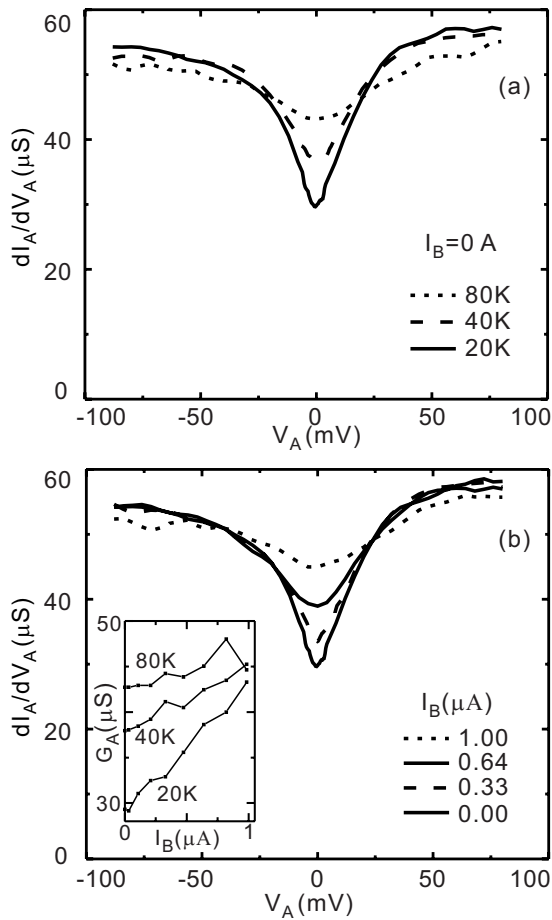}
\caption{ \label{fig3} Differential conductance
$dI_{A}/dV_{A}(V_{A})$ measured on SWNT $A$ (a) for different $T$
and (b) for different $I_{B}$ through SWNT $B$ at 20~K. The inset
in (b) shows $G_{A}$ for $V_{A}=0$ as a function of $I_{B}$.
$I_{B}=1\mu$A corresponds to $V_{B}=0.8$~V. The other points are
separated in bias by $\Delta V_{B}=0.1$~V.}
\end{figure}

Fig.~\ref{fig3}(a) shows the differential conductance
$dI_{A}/dV_{A}$ measured on tube $A$ as a function of $V_{A}$ for
different temperatures and with tube $B$ left floating. A clear
ZBA is observed, which becomes larger as temperature is decreased.
Such a ZBA has been observed many times in SWNTs
\cite{bockrath,yao,postma1}, and implies that a barrier lies along
the tube or at the interface with the electrodes.
Fig.~\ref{fig3}(b) shows $dI_{A}/dV_{A}(V_{A})$ when a current
$I_{B}$ is imposed to flow through the second tube. Interestingly,
the ZBA is progressively suppressed when $I_{B}$ is increased. We
note that the ZBA suppression depends only on the intensity of
$I_{B}$ and not on its sign. For these measurements, the sample
was biased such that no current flows from tube A to tube B
through the crossing point. In order to achieve this, first a
three-point measurement is carried out on tube $A$ under bias
$V_{A}$ to determine the potential $V_{A}^X$ at the crossing. The
voltage drops between the crossing and each electrode are recorded
as a function of $V_{A}$ and are found to be half of the bias
applied on the tube. In a second step, the three-point measurement
is carried out on tube $B$ to record the potential $V_{B}^X$ at
the crossing as a function of $V_{B}$. This time, the voltage
drops are very different on both sides of the tube reflecting the
large contact resistance at the $B1$ electrode. Finally, $I_{A}$
is measured as a function of $V_{A}$ for different $V_{B}$ where
voltages $V_{B1}$ and $V_{B2}$ applied on electrodes $B1$ and $B2$
are continually adjusted so that \mbox{$V_{A}^X$=$V_{B}^X$}, see
also Fig.~\ref{fig1}(b). Since most of $V_B$ drops at the bad
contact $B1$, we give instead of $V_B$ the current $I_B$ in
Fig.~\ref{fig3}(b) legend, which is measured while tube $A$ is
left floating. The differential conductance is obtained using
numerical differentiation.

We review now some possible explanations for the $I_B$-dependence
of the ZBA.  Let us first consider the effect of Joule heating.
Note that heating effects are generally disregarded in tunneling
experiments into individual tubes
\cite{bockrath,yao,postma1,bachtold2,tarkiainen,yi}. However, the
poor $B1$ contact releases significant heat in tube $B$. Part of
it flows through tube $A$, which may then change the temperature
sensitive $G_{A}$. Unfortunately, the temperature rise $\Delta T$
is difficult to estimate, already because little is known about
the thermal conductances of individual SWNTs and tube-tube
junctions. Nevertheless, a qualitative statement can be made.
Since $G_{A}(20$K$,0.6\mu$A$)\simeq G_{A}(40$K$,0$A$)$ and
$G_{A}(40$K$,0.6\mu$A$)\simeq G_{A}(80$K$,0$A$)$, the same heat
input 360~nW would give rise to temperature increases
20$\rightarrow$40~K and 40$\rightarrow$80~K. This would imply that
the thermal conductance decreases with $T$, which is very unlikely
in this $T$ range \cite{berber,zheng,pkim}. Hence thermal effects
alone cannot explain our observations. Another explanation might
be related to the capacitive coupling between the tubes. Metal
tubes can have an energy dependent conductance, which thus varies
with $V_{g}$  as in interference experiments \cite{kong,liang}.
Here the conductance $G_{A}$ is indeed observed to fluctuate with
$V_{g}$. One could thus argue that tube $B$ just acts as a gate.
However, the fluctuations with $V_{g}$, which are lower than
2.1~$\mu$S at 20~K and above, cannot account for the large
modulation of $G_{A}(I_{B})$. We conclude that another explanation
is needed to account for our experimental results.

Next we compare the data to Luttinger liquid predictions for two
crossed SWNTs with identical LL parameter $g$
\cite{komnik1,komnik2}.  Since the experiment is carried out at
zero tube-tube current, single-electron tunneling at the crossing
can be neglected,  and hence only tube-tube electrostatic coupling
and crossing-induced backscattering (CIB) processes need to be
taken into account. The importance of CIB processes due to the
tube deformation has been stressed in several previous
experimental \cite{janssen,hertel} and theoretical studies
\cite{rochefort,nardelli}.  Both are taken as local couplings
acting only at the crossing. Adopting the standard bosonization
formalism \cite{gogolin98}, for $g<1/5$, the most relevant part of
the density operator in tube $\alpha= A,B$ is \cite{egger}
\[
\rho_{\alpha}(x) \propto \cos[\sqrt{16\pi g}\
\varphi_{c+,\alpha}(x)],
\]
where $\varphi_{c+,\alpha}$ is the boson field describing charged low-energy
excitations (plasmons) of the SWNT.   Choosing spatial coordinates
such that $x=0$ corresponds to the crossing point, the Hamiltonian
$H=H_0+H_{\rm AB}+H_{\rm CIB}$ consists of the clean LL part,
 $H_0=\sum_\alpha H_{LL,\alpha}$,
a local tube-tube coupling $H_{\rm AB}= \lambda_0 \rho_A(0) \rho_B(0)$,
and the CIB part $H_{\rm CIB}= \lambda_1 \rho_A(0)
+ \lambda_2 \rho_B(0)$.  Standard
renormalization group (RG) analysis \cite{gogolin98}
yields the lowest-order flow equations
\begin{eqnarray}\label{rg1}
\frac{d \lambda_0}{d\ell} &=& (1-8g) \lambda_0 + 2 \lambda_1 \lambda_2,\\
\nonumber \frac{d \lambda_{1,2}}{d\ell} &=& (1-4g) \lambda_{1,2},
\end{eqnarray}
where $\ell$ is the usual flow parameter, $d\ell=-d\ln D$, i.e.,
one decreases the high-energy bandwidth cutoff $D$ and compensates
this decrease by adjusting the couplings. The initial coupling
constants $\lambda_{0,1,2}(0)$ could be accessed from microscopic
considerations, but here are only assumed to be non-zero.
Integration of Eq.~(\ref{rg1}) yields $\lambda_{1,2}(\ell)
=\lambda_{1,2}(0) \exp[(1-4g)\ell]$ and
\begin{eqnarray}\label{l2}
\lambda_0(\ell) &=& [\lambda_0(0)-2 \lambda_1(0)\lambda_2(0)]
e^{(1-8g) \ell}\\ \nonumber & +& 2\lambda_1(0)\lambda_2(0)
e^{(2-8g)\ell} .
\end{eqnarray}
Apparently,  at low energies (large $\ell)$, the RG flow is
completely dominated by $\lambda_0(\ell)$ due to the last term in
Eq.~(\ref{l2}). Ignoring the couplings $\lambda_{1,2}(\ell)$ at
such energy scales is then justified, and one can use the
single-channel model with only the $\lambda_0$ coupling of
Refs.~\cite{komnik1,komnik2}, taken at effective interaction
parameter $K_{\rm eff}= 4g-1/2$.  Taking $g=0.16$, this gives
$K_{\rm eff} = 0.14$.  For this argument, it is crucial that
$g<1/5$ and $\lambda_{1,2}(0)\neq 0$, for otherwise $\lambda_0$ is
irrelevant for all $g>1/8$. The CIB processes therefore drive the
electrostatic tube-tube coupling $\lambda_0$ to be the dominant
interaction in this crossed geometry.

The relevancy of the coupling $\lambda_0$ now generates a ZBA
which disappears when current flows in the second tube, in
agreement with experiments. For $K_{\rm eff}=1/4$ (corresponding
to $g=0.1875$), this can be made explicit by a simple analytical
solution of the resulting transport problem \cite{komnik2}. While
the exact solution can be obtained for any $K_{\rm eff}$ as well
\cite{egger00}, away from $K_{\rm eff}=1/4$ this solution is less
transparent and shows only slight differences.  We therefore focus
here on $K_{\rm eff}=1/4$, where the current through SWNTs $\alpha
= A,B$ is
\begin{equation}\label{e1}
I_{\alpha} = \frac{4e^{2}}{h} [V_{\alpha}-(U_{+} \pm U_{-})/\sqrt{2}],
\end{equation}
with $U_{\pm}$ obeying the self-consistency relations
\begin{equation}\label{e2}
eU_{\pm} = 2 kT_{B} \mathrm{Im}\Psi
\left(\frac{1}{2}+\frac{kT_{B}+i(eV_{\pm}-eU_{\pm})}{2\pi
kT}\right),
\end{equation}
with the digamma function $\Psi$, $V_{\pm}=(V_{A} \pm
V_{B})/\sqrt{2}$, and an effective coupling strength $T_{B}$,
which depends on the system parameters,
in particular on the initial couplings $\lambda_{0,1,2}(0)$.

\begin{figure}
\includegraphics{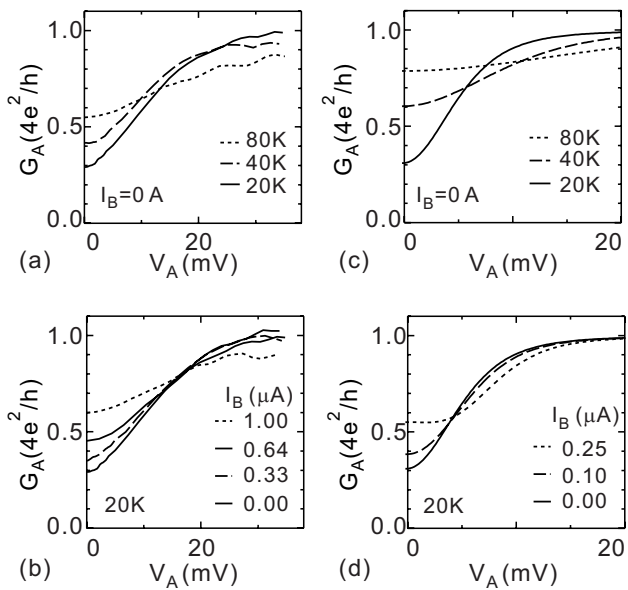}
\caption{ \label{fig4} Differential conductance
$dI_{A}/dV_{A}(V_{A})$ symmetrized and modified from Fig.~3 for
(a) different $T$ and (b) for different $I_{B}$. Theoretical
predictions for two interacting SWNTs are shown in (c) and (d).
The curves in (d) are obtained for constant biases $V_B$. The
corresponding currents $I_{B}$, which are calculated with Eqs.~(3)
and (4) for $V_{A}=0$, are given in the legend. }
\end{figure}

Figures \ref{fig4}(a-b) show modified $dI_{A}/dV_{A}(V_{A})$
curves of Fig.~\ref{fig3}. Indeed, Fig.~\ref{fig3} shows that the
high-bias differential conductance $dI_{A}/dV_{A}$ saturates at
$(17.9$~k$\Omega)^{-1}$ instead of $4e^{2}/h$, which is the
high-bias conductance predicted by Eqs.~(\ref{e1}) and (\ref{e2}).
We therefore argue that a resistance $R_{c}=11.4$~k$\Omega$ lies
in series with the $I_{B}$ dependent contribution of the
inter-tube coupling in order to obtain this $dI_{A}/dV_{A}$
saturation. The resistance $R_{c}$, presumably located at the
tube-electrode interfaces, is taken constant. This approximation
is quite good since the ZBA tends to disappear for large $I_{B}$,
leaving only a weak $1/R_{c}$ conductance modulation, see
Fig.~\ref{fig3}(b). Moreover, the conductance is known to change
only slightly with $T$ or $V$ in experiments on individual SWNTs
that are well contacted with contact resistance of the order
$10$~k$\Omega$ \cite{kong,liang}.

Figs.~\ref{fig4}(c,d) show the predicted $dI_{A}/dV_{A}(V_{A})$ curves
calculated from Eq.~(\ref{e1}) with (\ref{e2}).
The effective coupling $T_{B}$ is
set at $T_B= 11.6$~K to get agreement with the experimental value for
$G_{A}$ at $20$~K,  $I_{B}=0$ and $V_{A}=0$.
Despite the above-mentioned approximations, the agreement of theory and
experiment is quite good. After fixing $T_B$, no
parameter is tuned to calculate the
conductance variation with $V_{A}$, $T$ and $I_{B}$.  We note in passing that
Eqs.~(\ref{e1},\ref{e2}) predict the emergence of minima in
$dI_{A}/dV_{A}(V_{A})$ at large $I_{B}$, when $I_{B}\gtrsim$1
$\mu$A. These interesting features have not been observed though.
One probable cause could be the inelastic scattering on optical phonons, which
takes place at such large currents \cite{yao2}.  Scattering processes of
this kind are not included in Eqs.~(\ref{e1}) and (\ref{e2}).

We have presented experimental data
 for transport in crossed metallic single-wall carbon
nanotubes.  The results are in rather good agreement with a
theoretical analysis based on the Luttinger liquid model, and
cannot be rationalized by several alternative mechanisms.  We
therefore take this as new evidence for the Luttinger liquid
picture of SWNTs beyond tunneling experiments.

We thank B. Pla\c{c}ais, J.M. Berroir, N. Regnault and C.
Delalande for discussions.
LPMC is CNRS-UMR8551 associated to Paris 6 and 7. This
research has been supported by ACN programs (NN029),
the EU network DIENOW, the  DFG-SFB TR 12, and by the
Landesstiftung Baden--W\"urttemberg.

\end{document}